\documentclass[final,5p,times,twocolumn]{elsarticle}
\usepackage{amssymb}
\usepackage{graphicx}
\usepackage{dcolumn}
\usepackage{amsmath}

\begin{document}

\begin{frontmatter}

\title{Quantum topological transitions and spinons in metallic ferro- and antiferromagnets}

\author{V. Yu. Irkhin}

\ead{Valentin.Irkhin@imp.uran.ru}

%\pacs{75.10.Lp}{Band and itinerant models}
%\pacs{75.30.Mb, 71.28.+d}
\address
{M. N. Mikheev Institute of Metal Physics, 620108 Ekaterinburg, Russia
}
\begin{abstract}
An effective Hamiltonian describing fluctuation effects in the magnetic phases of the Hubbard model in terms of spinon excitations is derived. A comparison of spin-rotational Kotliar-Ruckenstein slave boson and Ribeiro-Wen dopon representations is performed. The quantum transition into the half-metallic ferromagnetic state with vanishing of spin-down Fermi surface is treated as the topological Lifshitz transition in the quasimomentum space.
The itinerant-localized magnetism transitions and Mott transition in antiferromagnetic state are considered in the  topological context. Related metal-insulator transitions in Heusler alloys are discussed.
\end{abstract}

%\pacs{71.27.+a, 75.10.Lp, 71.30.+h}

%\maketitle

\end{frontmatter}

\section{Introduction}

Exotic quantum phase transitions (QPT) in topological materials have
recently been extensively investigated \cite{Volovik}.
%change of Fermi surface topology
%book??.
Besides the simplest one-electron Lifshitz transitions, there is a more
complicated quantum scenario: vanishing of quasiparticle residue,
accompanied by spin-charge separation, and occurrence of incoherent states
and spinon Fermi surface \cite{Wen}.

QPT are treated in both usual paramagnetic and (in the presence of
frustrations) exotic spin-liquid states \cite{Vojta}.
%spin-charge separation
However, concepts of exotic QPT can be applied to ferromagnetic (FM) and
antiferromagnetic (AFM) phases too. Here belong also problems of magnetism
in high-$T_c$ cuprates \cite{Wen} and Kondo lattices \cite{Vojta,I17}.
%?? Weyl points

An important class of ferromagnets are half-metallic ferromagnets (HMF)
which possess unusual electronic properties connected with the presence of
states with only one spin projection at the Fermi surface and by energy gap
for another spin projection \cite{RMP}. In such a situation, incoherent
(non-quasiparticle) contributions play an important role. Vanishing of the
partial Fermi surface in the HMF state can be treated as a topological
transition in the quasimomentum space; a similar transition can be
considered in an antiferromagnet. We shall demonstrate that these systems
have quite non-trivial topological properties from the microscopic point of
view (within the many-electron models). To this end we develop the concept
of spinons in the magnetic phases by using the slave-particle
representations.

\section{Slave-particle representations}

We study the problem both in the Hubbard and s-d exchange (Kondo lattice)
models. The Hamiltonian of the former model reads
\begin{equation}  \label{eq:original_H}
\mathcal{H}_{\mathrm{H}} = \sum_{ij\sigma}
t_{ij}c^\dag_{i\sigma}c^{}_{j\sigma}+U\sum_i n_{i\uparrow}n_{i\downarrow},
\end{equation}
where $c^{\dagger}_{i\sigma}$ are electron creation operators.

First we treat the strong correlation limit. %To describe spinons,
Here it is convenient to use auxiliary (``slave'') boson and fermion
representations. Anderson's \cite{633a} representation exploiting the idea
of the separation of the spin and charge degrees of freedom of electron ($%
\sigma = \pm1$) reads
\begin{equation}
c_{i\sigma }=X_i(0,\sigma )+\sigma X_i(-\sigma,2)=e_i^{\dagger }f_{i\sigma
}+\sigma d_i f^{\dagger }_{i-\sigma }.  \label{eq:6.131}
\end{equation}
where $X$ are the Hubbard operators, $f_{i\sigma }$ are the annihilation
operators for neutral fermions (spinons), and $e_i$, $d_i$ for charged
spinless bosons (holons and doublons). For large $U$ and for hole doping
(where electron concentration $n<1$), we have to retain only the term with
holon operators. The choice of the Fermi statistics for spinons and Bose one
for holons is not unique and depends on the physical picture (e.g., presence
or absence of magnetic ordering, see also \cite{Kane}).

%??Alternative slave fermion - \cite{Kane} Weng - only quasiclassical??

%\begin{equation}
%c_{i\sigma }=X_i(0,\sigma )=f_i^{\dagger }b_{i\sigma }
% \label{eq:6.131c}
%\end{equation}

%slave fermion Kane -too simple?? only zero approxination?? Sz?? like phonons  конденсат - гамильтонинан Нагаева перейти к s-d модели
%не может быть точным из-за квадр корней? но входит b^2

%Thus,  the spin operators in Eq.~(\ref{eq:I.7}) are presented as the bilinear form %for expressed through the
%\begin{equation}
%\mathbf{S}_i=\frac 12\sum_{\sigma \sigma ^{\prime }}a_{i\sigma
%}^{\dagger }\mbox{\boldmath$\sigma $}_{\sigma \sigma ^{\prime
%}}a_{i\sigma ^{\prime }}, \,
% \label{eq:O.1}
%\end{equation}
%in terms of Schwinger bosons ($a_{i\sigma }^{\dagger } = b_{i\sigma }^{\dagger }$) or fermionic spinons ($a_{i\sigma }^{\dagger } = f_{i\sigma}^{\dagger }$), $\mbox{\boldmath$\sigma $}$ being Pauli matrices.

%??Whereas Gutzwiller method describes confinement Fermi-liquid state, but a more accurate consideration in terms of X-operators is required

A more complicated representation was proposed by Kotliar and Ruckenstein~%
\cite{Kotliar86}.
%introduces the slave boson operators $e_{i},\,p_{i\sigma },\,d_{i}$, so that
We use a rotationally invariant version \cite{Li,Fresard:1992a,Fresard1}.
This is suitable for magnetically ordered phases to take into account spin
fluctuation corrections and non-quasiparticle states treated in Refs. \cite%
{IK90,RMP}. We have
\begin{equation}
c_{i\sigma }=\sum_{\sigma ^{\prime }}f_{i\sigma ^{\prime }}z_{i\sigma
^{\prime }\sigma },~\hat{z}_{i}=e_{i}^{\dag }\hat{L}_{i}M_{i}\hat{R}_{i}\hat{%
p}_{i}+\widehat{\tilde{p}}_{i}^{\dag }\hat{L}_{i}M_{i}\hat{R}_{\hat{\imath}%
}d_{i}  \label{zz}
\end{equation}%
where
\begin{eqnarray}
\hat{L}_{i} &=&[(1-d_{i}^{\dag }d_{i})\sigma _{0}-2\widehat{p}_{i}^{\dag }%
\widehat{p}_{i}]^{-1/2}  \label{1z1} \\
\hat{R}_{i} &=&[(1-e_{i}^{\dag }e_{i})\sigma _{0}-2\widehat{\tilde{p}}%
_{i}^{\dag }\widehat{\tilde{p}}_{i}]^{-1/2}  \label{1z2} \\
M_{i} &=&(1+e_{i}^{\dag }e_{i}^{{}}+\sum_{\mu =0}^{3}p_{i\mu }^{\dag
}p_{i\mu }^{{}}+d_{i}^{\dag }d_{i}^{{}})^{1/2}.  \label{1z}
\end{eqnarray}%
The additional square-root factors in (\ref{1z1})-(\ref{1z}) can be treated
in spirit of mean-field approximation. In particular, the factor $M$ (missed
in earlier work \cite{Li}) is equal to $\sqrt{2}$ due to sum rule (\ref{sum}%
) and enables one to obtain an agreement with the small-$U$ limit and with
the FM case.
%where $f_{i\sigma },f_{i\sigma }^{\dag }$ are the  Fermi (spinon) operators, $e_{i},\,d_{i}$ the Bose operators.
The scalar and vector bosons $p_{i0}$ and $\mathbf{p}_{i}$ are introduced as
\begin{equation}
\hat{p}_{i}=\frac{1}{2}(p_{i0}\sigma _{0}+\mathbf{p}_{i}\mbox{\boldmath$%
\sigma $})
\end{equation}%
with $\mbox{\boldmath$\sigma $}$ being Pauli matrices and $\hat{\tilde{p}}%
_{i}=(1/2)(p_{i0}\sigma _{0}-\mathbf{p}_{i}\mbox{\boldmath$\sigma $})$ the
time reverse of operator $\hat{p}_{i}$. The constraints read
\begin{equation}
e_{i}^{\dag }e_{i}^{{}}+\sum_{\mu =0}^{3}p_{i\mu }^{\dag }p_{i\mu
}^{{}}+d_{i}^{\dag }d_{i}^{{}}=1,  \label{sum}
\end{equation}%
\begin{equation}
2d_{i}^{\dag }d_{i}^{{}}+\sum_{\mu =0}^{3}p_{i\mu }^{\dag }p_{i\mu
}^{{}}=\sum_{\sigma }c_{i\sigma }^{\dag }c_{i\sigma }^{{}}.
\label{eq:lmb-constraint}
\end{equation}

% at arbitrary $U$.

%which were earlier treated in the many-electron representation of Hubbard's operators (cf. Ref.\cite{338}).
Eq.(\ref{zz}) can be simplified in the case of magnetic ordering near
half-filling (the electron concentration $n<1$) where, in the mean-field
approach, $p_{0}=p^{z}=p\simeq 1/\sqrt{2}$ in the FM state (and
correspondingly in the AFM state in the local coordinate system), $e\simeq
\langle e\rangle =(1-n)^{1/2}$. Probably, this simplification in the local
coordinate system can work also in the systems with strong spin fluctuations
and short-range order (e.g., in the singlet RVB state), but not in the usual
paramagnetic state.%
%\begin{equation*}
%2\widehat{p}_{i}^{\dag }\widehat{p}_{i}=\frac{1}{2}(p_{i0}^{2}\sigma _{0}+(%
%\mathbf{S}_{i}\mbox{\boldmath$\sigma $}))
%\end{equation*}

%Omitting the term with doubles $d$??? and
%and small hole concentration??
We work with the projected electron operator
\begin{equation}
\tilde{c}_{i\sigma }=X_{i}(0\sigma )={c}_{i\sigma }(1-n_{i-\sigma }).
\end{equation}%
We perform the transformation by using the sum rule (\ref{sum}) and taking
into account the eigenvalues of the denominators in (\ref{1z1}), (\ref{1z2})
(cf. Ref. \cite{Fresard1}). Further on, we express again the numerator in
terms of the Pauli matrices using explicitly their matrix elements.
Neglecting the terms proportional to holon operators we derive%
\begin{equation*}
\tilde{c}_{i\uparrow }=\frac{1}{\sqrt{2}}f_{i\uparrow }(p_{i0}+p_{iz}),~%
\tilde{c}_{i\downarrow }=f_{i\uparrow }p_{i}^{-}
\end{equation*}
%are given by $\pm 1$ in our representation,
Since the terms proportional to $f_{i\downarrow }$ and $p_{i}^{+}$ are small
(note that in the magnetic ordering case $p_{i}^{+}$ is not related to spin
operators, see below Eq.(\ref{SS1})) we can restore rotational invariance and write
down approximately
\begin{equation}
\tilde{c}_{i\sigma }=\sqrt{2}\sum_{\sigma ^{\prime }}\hat{p}_{i\sigma
^{\prime }\sigma }f_{i\sigma ^{\prime }}=\frac{1}{\sqrt{2}}\sum_{\sigma
^{\prime }}f_{i\sigma ^{\prime }}[\delta _{\sigma \sigma ^{\prime }}p_{i0}+(%
\mathbf{p}_{i}\mbox{\boldmath$\sigma $}_{\sigma ^{\prime }\sigma })].
\label{eq:I.88}
\end{equation}
This representation satisfies exactly commutation relations for Hubbard's
operators.

We see that the factor $e^{\dag }$ in the numerator of (\ref{zz}) is
canceled and the system has only spin degrees of freedom. On the other hand,
in the paramagnetic case (in particular, in the problem of the
metal-insulator transition), the charge fluctuations connected with $e$ are
important (see, e.g., \cite{Castellani,Fr}). 
%Thus the above transformation breaks when f-states become itinerant. 
The situation is similar to Weng's
consideration of confinement in cuprates \cite{Weng}. According to this,
%holons are confined and vice versa
%Weng: in saturated afm holons are confined while spinons are deconfined, in non-saturated  - motion of holes, damping of spin waves holes disappear from KR representation, in PM phase - only bosons - see lavagna??
in the underdoped regime the antiferromagnetic and superconducting phases
are dual: in the former, holons are confined while spinons are deconfined,
and vice versa, and the gauge field, radiated by the holons (spinons),
interacts with spinons (holons) through minimal coupling.
%A U(1) gauge field Ah  (As ), radiated by the holons (spinons), interacts with spinons (holons) through minimal coupling \cite{Weng}.
%in non-saturated  - motion of holes, damping of spin waves holes disappear from KR representation, in PM phase - only bosons - see lavagna??

%\tilde{p} differs from p by change of spin projections S to -S we change phase b = -1 to correct physics \overrightarrow
Defining spinor operators $\tilde{c}_{i}^{{}}=(\tilde{c}_{i\uparrow }^{{}},\,%
\tilde{c}_{i\downarrow }^{{}})$ etc. we derive the beautiful representation
\begin{equation}
\tilde{c}={f}\hat{p}  \label{eq:1.88}
\end{equation}%
The vector gapless boson $\mathbf{p}$
%play a role unlike vector gauge bosons - Ref. - material field
restores the rotational invariance and describes spin degrees of freedom
since
\begin{equation}
\mathbf{S}=\frac{1}{2}\sum_{{}}\mbox{\boldmath$\sigma $}_{\sigma \sigma
^{\prime }}p_{\sigma \sigma _{1}}^{\dag }p_{\sigma _{2}\sigma ^{\prime }}=%
\frac{1}{2}(p_{0}^{\dag }\overline{\mathbf{p}}+\overline{\mathbf{p}}^{\dag
}p_{0}-i[\overline{\mathbf{p}}^{\dag }\times \overline{\mathbf{p}}])
\label{SS}
\end{equation}%
with $\overline{\mathbf{p}}=(p^{x},-p^{y},p^{z})$. The corresponding
spectrum $\omega _{\mathbf{q}}$ is determined by effective intersite
exchange interaction or by additional Heisenberg interaction
\begin{equation*}
\mathcal{H}_{d}=\sum_{\mathbf{q}}J_{\mathbf{q}}\mathbf{S}_{-\mathbf{q}}%
\mathbf{S}_{\mathbf{q}}
\end{equation*}%
in the $t-J$ model and has essentially spin-wave form. In the FM case
%($p_0\simeq p^z \simeq 2^{-1/2}$)
we obtain $S_{i}^{+}\simeq p_{i}^{-}$, so that
\begin{equation}
\mathcal{H}_{d}=\sum_{\mathbf{q}}\omega _{\mathbf{q}}(p_{\mathbf{q%
}}^{-})^{\dag }p_{\mathbf{q}}^{-}+\mathrm{const},\,\omega _{\mathbf{q}}=J_{%
\mathbf{q}}-J_{0}.
\label{SS1}
\end{equation}%
with $p^\pm=2^{-1/2}(p^x \pm ip^y)$.
%
%Heisenberg Hamiltonian takes the standard spin-wave form
It should be stressed that the vector product in (\ref{SS}) has to be
retained to derive this result (otherwise the bosons $p$ and $p^{\dag }$ are
mixed), unlike the consideration in Ref.\cite{Fresard:1992a}.

%+gauge fields - Tremblay

%??A field U(1) an analogy with FL* with magnetic order??

%HFM state is stable - Sachdev introduces additional field
%Fm - ffm difference - non-uniform in space, true gap

%Unlike the X-operator approach narrow band  large-N expansion

To describe doped cuprates, also a representation of the Fermi dopons $%
d^\dag_{i\sigma}$ was proposed \cite{Ribeiro,654,Scr},
\begin{equation}
\tilde{c}_{i -\sigma }=-\frac{\sigma}{\sqrt{2}}\sum_{\sigma ^{\prime
}}d^{\dagger}_{i\sigma ^{\prime }}(1-n_{i-\sigma ^{\prime }}) [S\delta
_{\sigma \sigma ^{\prime }}-(\mathbf{S}_i\mbox{\boldmath$\sigma $}_{\sigma
^{\prime} \sigma} )].  \label{eq:I.78}
\end{equation}
where $\sigma=\pm$, $n_{i\sigma}=d^{\dagger}_{i\sigma}d_{i\sigma}$, and both
Fermi spinon (Abrikosov) and Schwinger boson representations can be used for
localized $S=1/2$ spins.
%??We see that the angular momentum  construction is the same as in (\ref{eq:I.88}) (see also Refs.\cite{654,Scr}).
In the magnetic case the structure of (\ref{eq:I.78}) is identical to the
spin-rotation invariant representation (\ref{zz}), except for the factor of $%
M\simeq \sqrt{2}$. To incorporate such a correction in the dopon
representation, we note that one can include into (\ref{eq:I.78}) an
additional square root factor in terms of spinon and dopon operators,
\begin{eqnarray}
\tilde{c}_{i -\sigma }=-\frac {\sigma}{\sqrt{2}}\sum_{\sigma ^{\prime }}
[1+\sum_{\sigma^{\prime \prime }} (f^\dag_{i\sigma^{\prime \prime
}}f^{}_{i\sigma^{\prime \prime }} - d^\dag_{i\sigma^{\prime \prime
}}d^{}_{i\sigma^{\prime \prime }})]^{1/2}  \notag \\
\times d^{\dagger}_{i\sigma ^{\prime }}(1-n_{i-\sigma ^{\prime }}) [S\delta
_{\sigma \sigma ^{\prime }}-(\mathbf{S}_i\mbox{\boldmath$\sigma $}_{\sigma
^{\prime }\sigma} )].  \label{eq:I.78a}
\end{eqnarray}
This factor has also eigenvalue of 1 on the physical space, but in the
mean-field approach its average yields $\simeq \sqrt{2}$, which cancels the
corresponding prefactor in (\ref{eq:I.78}).

The discrepancy is also removed in the classical limit of the s-d model
(with arbitrary large $S$ in (\ref{eq:I.78})), cf. Ref.\cite{Scr}.
%This difficulty may be similar to spurious factor of 2 in the exponent of the Kondo temperature in the Anderson lattice model (see  Refs.\cite{Rice,Moller} for the Gutzwiller approximation), a correct result for the effective hybridization being  restored in the $1/N$-expansion.
%($N-1 \rightarrow N$ in the effective hybridization).
%??Note that an effective spinon-dopon  hybridization is also introduced in Ref.\cite{Ribeiro}.
%??- Appendix in Ribeiro-Wen - 1/N corrects??
It is interesting to note that spinons $f_{i\sigma}$ and dopons $d_{i\sigma}$
are interchanged in the representations (\ref{eq:I.88}) and (\ref{eq:I.78}).

The representation (\ref{eq:I.78a}) can connect physics of the Kondo
lattices and cuprates. Such an approach was developed also in Ref.\cite{Punk}
to describe the formation of spin liquid state in terms of frustrations in
localized-spin subsystems, the Schwinger boson representation being used for
the latter. One can see that according to (\ref{eq:I.78a}) the distribution
functions of dopons and physical electrons are simply related as
\begin{eqnarray}
\langle\tilde{c}^\dag_{\mathbf{k}\sigma}\tilde{c}_{\mathbf{k}\sigma}\rangle
=- (\frac12+\sigma \langle S^z \rangle)^2 \langle d^\dag_{-\mathbf{k}%
\sigma}d_{-\mathbf{k}\sigma}\rangle  \notag \\
+ \mathrm{smooth \, function \, of}\, \mathbf{k.}  \label{dd1}
\end{eqnarray}
which differs from the consideration in Ref. \cite{Punk} by the absence of
the factor 1/2. The result (\ref{dd1}) is in agreement with those for the
saturated FM case (Nagaoka state) where the Green's function residue for
spin-up states equals unity, and spin-down states demonstrate purely
non-quasiparticle (incoherent) behavior with zero residue \cite{IK90}.
%which has the same quantum numbers as the electron

%Taking into account the condition $\sum_{\sigma} f^\dag_{i\sigma}f^{}_{i\sigma}=1$,
At small doping the operator (\ref{eq:I.78a}) can be rewritten as
\begin{equation}
\tilde{c}_{i\sigma}=(d_{i\downarrow}^\dag f_{i\uparrow}^\dag
-d_{i\uparrow}^\dag f_{i\downarrow}^\dag )f_{i\sigma}
\end{equation}
Introducing the Bose holon operator $e_i=f_{i\uparrow}d_{i\downarrow}-f_{i%
\downarrow}d_{i\uparrow}$ we return to Anderson's representation (\ref%
{eq:6.131}). We see again that the absence of the factor 2$^{-1/2}$ is
needed.

%??On the other hand, in the magnetic case it convenient to  use the Schwinger rather than the spinon representation
%???????for the dopons $d_{i\sigma}$ (see (\ref{eq:O.1})).
%??The supersymmetry approach  \cite{Lavagna} mixes the fermionic and the bosonic representation of the spin following the standard rules of superalgebra.

\section{Electron Green's functions in the HMF state}

Consider the electron Green's functions for a saturated ferromagnet in the
representation (\ref{eq:I.88}). The spin-up states propagate freely on the
background of the FM ordering, but the situation is quite non-trivial for
spin-down states. Spin-down state is a complex of spinon $f^\dag_\uparrow$
and boson $(p^-)^{\dag}$ (generally speaking, coupled by gauge field):
% we obtain
\begin{equation}
G_{\mathbf{k}\downarrow }(E)=\sum_{\mathbf{q}\mathbf{q^{\prime }}}\langle
\langle p^-_{\mathbf{q}}f_{\mathbf{k}-\mathbf{q}\uparrow}| f^{\dagger}_{%
\mathbf{k-q^{\prime }}\uparrow}(p^-_{\mathbf{q^{\prime }}})^{\dag} \rangle
\rangle_E  \label{eq:I.778}
\end{equation}
The simplest decoupling in the equation of motion for the Green's function
in the right-hand side of (\ref{eq:I.778}) yields
\begin{equation}
G_{\mathbf{k}\downarrow }^0(E)=\sum_{\mathbf{q}}\frac{1+N(\omega_{\mathbf{q}%
})-n(t_{\mathbf{k}+\mathbf{q}})}{E-t _{\mathbf{k}-\mathbf{q}}-\omega _{%
\mathbf{q}}}  \label{eq:I.779}
\end{equation}
with $N(\omega)$ and $n(E)$ being the Bose and Fermi functions. A more
advanced decoupling yields the result
\begin{equation}
G_{\mathbf{k}\downarrow }(E)=\left\{ E-t_{\mathbf{k}}+\left[ G_{\mathbf{k}%
\downarrow }^0(E)\right] ^{-1}\right\} ^{-1}  \label{eq:I.776}
\end{equation}
These results were obtained earlier by treating the Hubbard ferromagnet in
the many-electron representation of Hubbard's operators \cite{FTT,IK90}, the
analogy with Anderson's spinons (which are also described by the Green's
function with zero residue) being mentioned. The Green's function (\ref%
{eq:I.779}) has a purely non-quasiparticle nature. Because of the weak $%
\mathbf{k}$-dependence of the corresponding distribution function the
non-quasiparticle (incoherent) states possess a small mobility and do not
carry current.

At small doping $1-n$, the Green's function (\ref{eq:I.776}) has no poles
above the Fermi level, so that the above conclusions are not changed.
However, with increasing $1-n$, it acquires a spin-polaron pole above $E_F$,
and the saturated ferromagnetism is destroyed.

%??other scenarios of crossover include spin-wave instabilities??

The description of the transition to the saturated state (vanishing of the
quasiparticle residue) is similar to that of the Mott transition in the
paramagnetic Hubbard model \cite{Senthil}. Thus the situation is somewhat
analogous to a partial Mott transition in the spin-down subband, see the
discussion of the orbital-selective Mott transition in the review \cite%
{Vojta}.

%Spin-down electrons occur with the pole at the Fermi level and are absent in the saturated ferromagnet.

%orbital-selective Mott transition - both in afm fm and pm??
%Luttinger surface and Kondo pole!!!

%??We discuss criteria for the existence of topological transitions in the antiferromagnetic phase - Vojta1

%not simple Lifshitz - vanishing of residue

%Role of Van Hove singularities which have topological origin too??

%the topological transition
%Is similar to crossover - temperature dependences of transport properties change, resistivity since one-magnon spin-flip processes are forbidden

Now we pass to the case of finite coupling which can be treated both in the
Hubbard and s-d models \cite{Edwards:1973,IK90}). The Hamiltonian of the
latter model reads
\begin{equation}
\mathcal{H}= \sum_{\mathbf{k}\sigma }t_{\mathbf{k}}c_{\mathbf{k}\sigma
}^{\dagger }c_{\mathbf{k}\sigma }+\mathcal{H}_d -I\sum_{i\sigma \sigma
^{\prime }}(\mathbf{S}_i\mbox{\boldmath$\sigma $}_{\sigma \sigma ^{\prime
}})c_{i\sigma }^{\dagger }c_{i\sigma ^{\prime }},  \label{eq:G.2}
\end{equation}
Note that magnetic ordering in this model as a rule does not vanish owing to
existence of local momemts and intersite exchange interactions (except for
the Kondo effect situation which can occur at $I<0$). Thus we have only one
phase transition -- from saturated to non-saturated state. Consistent
solution for $I>0$ in the HMF state yields \cite{353,I15} %
%\begin{equation}
%\Sigma _{\mathbf{k}\uparrow }(E)=2I^2S
%\sum_{\mathbf{q}}\frac{N_{\mathbf{q}}+n_{\mathbf{k}+\mathbf{q}\downarrow
%}}{E-t_{\mathbf{k}+\mathbf{q}\downarrow }+\omega _{\mathbf{q}}}, \,
%\Sigma _{\mathbf{k}\downarrow }(E)= 2I^2S \sum_{\mathbf{q}}
%\frac{1+N_{\mathbf{q}}-n_{\mathbf{k}-\mathbf{q}\uparrow
%}}{E-t_{\mathbf{k}-\mathbf{q}\uparrow }-\omega _{\mathbf{q}}},
% \label{eq:G.34}
%\end{equation}
\begin{equation}
G_{\mathbf{k\uparrow }}^{{}}(E)=(E-t_{\mathbf{k}}+IS)^{-1}  \label{up}
\end{equation}
\begin{eqnarray}
G_{\mathbf{k\downarrow }}^{{}}(E) &=&\left( \ E-t_{\mathbf{k}}+IS-\frac{2IS}{%
1-IR_{\mathbf{k\uparrow }}(E)}\right) ^{-1},  \label{down} \\
R_{\mathbf{k\uparrow }}(E) &=&\sum_{\mathbf{q}}(1-n_{\mathbf{k-q\uparrow }%
})G_{\mathbf{k-q\uparrow }}(E-\omega _{\mathbf{q}})  \notag  \label{I>0}
\end{eqnarray}
%with $t_{\mathbf{k\sigma }}=t_{\mathbf{k}}\mp IS$.
The result for the Hubbard model differs from (\ref{I>0}) by the replacement
$I\rightarrow U$ \cite{IK90}. The incoherent states occur above the Fermi
level.

%??NQP states were recently observed experimentally

For $I<0$, $G_{\mathbf{k\downarrow }}^{{}}(E)$ has the same form, and we can
write down an approximate solution
\begin{eqnarray}
G_{\mathbf{k\uparrow }}^{{}}(E) &=&\left( \ E-t_{\mathbf{k}}-IS+\frac{2IS}{%
1+IR_{\mathbf{k\downarrow }}(E)}\right) ^{-1},~  \label{up1} \\
R_{\mathbf{k\downarrow }}(E) &=&\sum_{\mathbf{q}}n_{\mathbf{k-q\downarrow }%
}G_{\mathbf{k-q\downarrow }}(E+\omega _{\mathbf{q}})  \notag
\end{eqnarray}
so that the incoherent states occur below the Fermi level. The cases $I>0$
and $I<0$ are not simply related by the particle-hole transformation because
of the Kondo divergence of the denominator in (\ref{up1}) in the latter
case, which indicates a quantum phase transition with increasing electron
concentration or $|I|$. A consistent treatment of this case in the large-$%
|I| $ case where the incoherent states predominate is given in Ref.\cite{I15}%
. For $I\rightarrow -\infty$ we have
\begin{equation}
G_{\mathbf{k}}^{\downarrow }\left( E\right) =\frac{2S}{2S+1}(\epsilon -t_{%
\mathbf{k}}^{\ast })^{-1}  \label{I-}
\end{equation}%
\begin{eqnarray}
G_{\mathbf{k}}^{\uparrow }\left( E\right) &=&\frac{2S}{2S+1}\left[ \epsilon
-t_{\mathbf{k}}^{\ast }+\frac{2S}{R_{\mathbf{k}}^{\ast }(\epsilon )}\right]
^{-1},  \label{nqpl} \\
R_{\mathbf{k}}^{\ast }(\epsilon ) &=&\sum_{\mathbf{q}}\frac{n(t_{\mathbf{k-q}%
}^{\ast })}{\epsilon -t_{\mathbf{k-q}}^{\ast }+\omega _{\mathbf{q}}}
\end{eqnarray}%
with $\epsilon =E-I(S+1),t_{\mathbf{k}}^{\ast }=[2S/(2S+1)]t_{\mathbf{k}}$.
For $S=1/2$ this case is equivalent to the Hubbard model with the
replacement $t_{\mathbf{k}} \rightarrow t_{\mathbf{k}}/2$.

%what are spinons???

\section{The case of antiferromagnets and discussion}

Now we consider the antiferromagnetic case. The slave boson representation
in the local coordinate system (cf. \cite{Fresard1}) yields the same spinon
form (\ref{eq:I.88}).

Mott transition in antiferromagnets usually goes in two steps: first from
the AFM insulator to AFM metal (the energy gap between AFM subbands
vanishes), and then from AFM metal to paramagnetic metal \cite{Timirgazin}.
Thus the situation is similar to half-metallic ferromagnetism, but even more
distinct: here we have the true insulator gap for both spin directions. In
the situation of doping, we have again the transition between two types of
AFM metals which can be denoted as saturated (localized-moment) and
non-saturated (itinerant). %In the limit of weak correlations
These types correspond to classification of Ref.\cite{Sokol}: type A (when
the Fermi surface does not cross the magnetic zone boundary) and type B
(when the Fermi surface crosses the magnetic zone boundary). Although in the
weak-coupling case this is a simply geometrical difference, in the case of
strong correlations (in particular, in the Hubbard model with large $U$) the
gap has essentially a Mott-Hubbard nature. Note that, unlike FM case, the
band splitting in the saturated AFM state can be very small.

Delocalization transitions from local-moment to itinerant magnetism take
place also in the Kondo systems \cite{Vojta}.
%below??intermediate non-saturated phase - non-integer (partially suppressed) magnetic moments,  with magnetic entropy being smaller than $R \ln2$, as well as in the Kondo lattices.
Such a delocalization phase diagram can be considered for both ferro- and
antiferromagnetic Kondo lattices \cite{Hoshino}.

The numerical calculations in the large-$U$ Hubbard model within the
slave-boson approach were performed in Ref.\cite{Igoshev}. For the
two-dimensional (2D) square lattice with finite next-nearest-neighbor
transfer integral $t^{\prime }$ they demonstrate the first-order transition
from HMF to paramagnetic (PM) state with increasing hole doping and
second-order transition from saturated AFM to PM state with electron doping.
At the same time, for the 3D cubic lattice the intermediate non-saturated FM
phase occurs. The difference is due to strong Van Hove singularities in the
electron spectrum of the square lattice which favor saturated state; this
circumstance has also a topological nature.

%partial frustration

%In particular, frustrations caused by doping induce disordered spin-liquid-like state \cite{Wen,Weng}.
According to Ref \cite{Coleman01}, a breakdown in the composite nature of
the heavy electron can take place at the quantum critical point between AFM
and Kondo heavy Fermi-liquid states.

The paramagnetic Mott insulator state can be related to deconfinement of
spinons and holons \cite{Senthil}. The deconfinement magnetically ordered
spin-density-wave state SDW$^*$ with a reduced moment can be also considered
\cite{Sachdev,Vojta}.

%%ultimately lead to the same results in the undoped antiferromagnet, but the latter is far more convenient in the doped antiferromagnet. Also, the latter
%This approach enables one to describe the state with small  Fermi surface without symmetry breaking.

%Topologically protected Z2
%Fermi liquid large Fermi surface

Unlike 2D systems, where monopoles prevent deconfinement, in the 3D systems
the U(1) gauge theory admits a deconfined phase where the spinons
potentially survive as good excitations. This deconfined phase has a Fermi
surface of spinons coupled minimally to a gapless ``photon'' (U(1) gauge
field), whereas monopoles are gapped \cite{Sachdev}.
%This is the advocated U(1) FL* phase.
Transitions to SDW* in 3D situation are more probable too.

The situation with deconfinement can be treated in different ways. According
to \cite{Weng1}, there is no true spin-charge separation in the ordered
phases, but the spin-charge separation (or deconfinement) can be treated as
a driving force in the unconventional phase transitions.

%To summarize, a
%crucial aspect of this SU(2) gauge theory description is
%to have chargons whose dispersions are identical (at the
%mean-field level) to the excitations of a spin-density wave
%states, despite the theory having no long-range order or
%broken symmetries (i.e. the symmetry is restored by the
%fluctuations of the spinon fields).
%Pseudogap and Fermi surface topology in the two-dimensional Hubbard model +Volovik

%2d - monopoles prevent topological phases

%approximate zeros of Green's function!!

%вихри и ежи в ФМ?? состояния со спином вниз заперты
%ферроны и квазиосцилляторы в афм -- струны потенциал возвращает
%связан ли запрет на движение с топологическим исчезновением ПФ?
% s-d модель - естественно Хиггс+dopons 

%Kane-Lee possible spin liquid state\cite{Senthil}

%At the same time, in afm hot spots

Since one-magnon scattering processes are forbidden in the half-metallic
ferromagnets and saturated antiferromagnets, usual and half-metallic systems
demonstrate different temperature dependences of electronic and magnetic
properties and characteristics, e.g., of resistivity, spin-wave damping etc.
\cite{UFN,RMP}.
%including crossovers in resistivity etc. at $T \sim T^*$ with $T^*$ being determined by magnetic spin splitting \cite{RMP}.

Let us discuss some experimental examples of topological transitions in
magnetic systems. The compound Co$_{2}$TiSn, which is supposed to be HMF,
demonstrates transition from semiconducting to metallic state as the system
undergoes the paramagnetic to ferromagnetic transformation \cite{Co2TiSn}.
CoFeTiSn shows the same feature;
%CoFeTiSn is similar to the compound Fe$_2$VAl  which has pseudogap in the paramagnetic phase.
on the other hand, CoFeVGa demonstrates a semiconducting behaviour down to
90~K, below which it shows a window of metallic region and
antiferromagnetism \cite{CoFeTiSn}. First principle calculations yield
nearly half-metallic electronic structure for CoFeTiSn and CoFeVGa (see
discussion in Ref.\cite{CoFeTiSn}), which may indicate topological
instabilities of the Fermi surface. Note that the saturated HMF state is
most stable in the mean-field theory of Kondo lattices \cite{608,I17};
inclusion of additional field is required to obtain more exotic states \cite%
{Sachdev}. In Ref. \cite{DMFT} the HMF state was rediscovered in the Kondo
model within the framework of the DMFT method and called \textquotedblleft
spin-selective Kondo insulator\textquotedblright\ (remember again the
analogy with the orbital-selective Mott state \cite{Vojta}).

%first order transition at large t'<0??+Kim

%that the majority electrons are metallic in CoFeTiSn, while the Fermi level touches the top of the minority valence band, and the material is described as a quasi-half metallic ferrimagnet with high degree of spin polarization [13]. CoFeVGa is also found to be a nearly half metal on the basis of first principle calculations [17].

The compound UNiSn turns out to be an antiferromagnet, although the band
structure calculations yield a HMF structure (see references in \cite{UFN}).
An unusual transition from metallic AFM state to small-gap semiconductor PM
state with increasing $T$ takes place at 47~K. Most simple explanation is
that the emergence of the sublattice magnetization results in a shift of the
Fermi level outside the energy gap, but the Kondo origin of the gap seems to
be more probable.
%[496] H.Fuiji, H.Kawanaka, T.Takabatake et al, J.Phys.Soc.Jpn 58, 3481 (1989) [497] M.Yethirai, R.A.Robinson, J.J.Rhyne, J.A.Gotaas and K.H.J.Bushow, J.Magn.Magn.Mat.79, 355 (1993)
Then the possible explanation of the metal-insulator transition observed is
that the AFM exchange interaction suppresses the Kondo order parameter $V
\sim \langle c^\dag f \rangle$ and, consequently, the gap \cite{UFN}. This
transition can be treated as a topological transition from large to small
Fermi surface \cite{Sachdev,Si}.

%\section{Acknowledgments}
The author is grateful to Yu.~N.~Skryabin and R.~Fresard for helpful
discussions. The research was carried out within the state assignment of
FASO of Russia (theme ``Quantum'' No. AAAA-A18-118020190095-4) with partial
support of the Russian Foundation for Basic Research (project No.
18-02-00739).
%and supported in part by the Russian Foundation for Basic Research (project no.16-02-00995).theme ``Flux'' No AAAA-A18-118020190112-8 and

\end{document}